\documentclass[useAMS,usenatbib]{mn2e}

\usepackage{graphicx}
\usepackage{natbib}
\usepackage{aas_macros}
\begin{document}

\title[DDO survey of MOST variables]{DDO spectroscopic
survey of MOST variable stars\thanks{Based on
spectroscopic data from the David Dunlap Observatory,
University of Toronto
}}

\author[Theodor Pribulla et al.] {Theodor Pribulla$^{1,2}$,
Slavek Rucinski$^1$, Rainer Kuschnig$^3$, 
\newauthor
Waldemar Og{\l}oza$^4$,
Bogumil Pilecki$^5$ \\
$^1$David Dunlap Observatory, University of Toronto,
P.O.~Box 360, Richmond Hill, Ontario, L4C~4Y6,
Canada\\
$^2$Astronomical Institute, Slovak Academy of Sciences,
059~60 Tatransk\'a Lomnica, Slovakia\\
$^3$Institut f\"{u}r Astronomie, Universit\"{a}t Wien,
T\"{u}rkenschanzstrasse 17, A-1180 Wien, Austria\\
$^4$Mt. Suhora Observatory of the Pedagogical University
ul.~Podchora\.{z}ych 2, 30-084 Cracow, Poland\\
$^5$Warsaw University Astronomical Observatory,
Al. Ujazdowskie 4, 00-478 Warszawa, Poland
}
\date{Accepted 0000 Month 00. Received 0000 Month 00;
in original form 2007 March 17}

\pagerange{\pageref{firstpage}--\pageref{lastpage}} \pubyear{2008}

\maketitle

\label{firstpage}

\begin{abstract}
A spectroscopic support survey of 103 objects observed by 
the MOST satellite is presented; 96 are variable stars with 83 of them 
being new MOST variable-star detections or stars with
variability types verified and/or modified on the basis of the MOST data.
Analysis of 241 medium-resolution spectra using the broadening-functions
formalism yielded radial velocities, projected rotational velocities
(for 31 targets for which it was possible) and spectral type estimates.
Seven new spectroscopic binaries 
were discovered; orbital solutions are given for two of
them (HD~73709, and GSC~0814-0323).
The visual binary HD~46180 was found to be composed of two close
binary stars (eclipsing and non-eclipsing one) very probably forming a
physical quadruple system.

\end{abstract}


\section{Introduction}
\label{intro}

MOST (Microvariability \& Oscillations of STars) is a microsatellite
dedicated to detection and characterisation of variability
of stars. It uses a 15-cm telescope which feeds a CCD photometer through
a single, custom, broadband optical filter (350 -- 700 nm).
The pre-launch characteristics of the mission are described
by \citet{WM2003} and the initial post-launch performance
by \citet{M2004}. Although the main goal of the mission was
to study pulsations of stars, continuous photometry
of selected fields led to discovery of many new variables of
different types. Several new variables were found among 
guide stars used to accurately point the telescope.
After the failure of the CCD used for the satellite stabilization
in January 2006, the number of guide stars utilized for this
purpose in the science CCD was increased to a few dozen per
field which led to detection of many variable stars. 
Observations in a single, wide-band
filter and small photometric amplitudes often
prevented correct determination of variability type
for newly discovered objects. The brightness range where spectroscopic
information is poor or absent is typically within $8 < V <11$.

\begin{table*}
\begin{scriptsize}
\caption{Objects observed within the DDO survey and their characteristics.
         \label{main.tab}}
\begin{center}
\begin{tabular}{llrrllrrrcl}
\hline
\hline
  HD/BD    &   GSC     & Cluster &  $V$ & Sp.    & Sp  & \# & RV    &$v \sin i$ & Multiplicity    & Note                     \\
           &           &         &      & publ.  & BF  &    &[km/s] & [km/s]    &                 &                          \\

 HD~965     & 4664-0318 &         &  9.09& A5p    & A5:    &  3 &  0.7 &     &      & roAp suspected, no apparent puls.   \\
 HD~6242    & 0022-0517 &         &  7.98& K2     & K7(G)  &  1 & 36.7 &     &      & LPV-known                           \\
 HD~8801    & 2817-2326 &         &  6.42& A7m    & F3:    &  3 &  3.4 &   58&      &                                     \\
 HD~9014    & 5276-0313 &         &  6.48& A9V    & F4     &  4 &      &     & SB2  & dScut-new                           \\
 HD~9289    & 5274-0240 &         &  9.71& A3     & A9     &  1 & 10.1 &     &      & roAp-known, spots-new               \\
 BD+37 410  & 2816-2068 & NGC752  &  9.94& F4III  & F5     &  2 &      &     & SB1  & eclBin-new                          \\
 HD~11720   & 2816-1892 & NGC752  &  8.09& K0III  & K0(G)  &  2 & 20.0 &     &      & RGpuls-new                          \\
 BD+36 364  & 2816-2276 & NGC752  & 10.44& F2III  & F4     &  8 &      &     & SB2  & gDor-new                            \\
 BD+37 431  & 2816-2273 & NGC752  &  9.85& F2III  & F4     &  4 &      &     & SB2  & gDor-new                            \\
 HD~11885   & 2816-0111 & NGC752  &  7.12& K1III  & K2(G)  &  2 & 10.7 &     &      & RGpuls-new                          \\
            & 2816-2061 & NGC752  & 10.44& F2V    & F4     &  1 &  4.5 &  130&      & gDor-new                            \\
            & 2816-2039 & NGC752  & 10.90& F0V    & F3     &  1 &  3.6 &   41&      & gDor-new                            \\
 BD+36 373  & 2816-0432 & NGC752  & 10.60& F2V    & F2     &  1 &  5.7 &   45&      & eclBin-new                          \\
 BD+37 440  & 2816-1973 & NGC752  &  9.13& F1V    & F2     &  3 &  1.9 &   39&      & gDor-new                            \\
 BD+37 444  & 2816-2176 & NGC752  &  9.62& F3V    & F3     &  8 &      &     & SB2  & gDor-new                            \\
 BD+37 445  & 2816-2225 & NGC752  & 10.70& F2III  & F7     &  1 &-55.1 &   45&      & var-new                             \\
 HD~12899   & 0636-0631 &         &  8.14& F0     & F4     &  4 & -6.7 &  141&      & dScut-new                           \\
            & 0636-0130 &         & 10.15&        & K0(G)  &  1 &  2.8 &     &      & LPV-new ?                                 \\
 BD+13 336  & 0636-0976 &         &  9.30& A5     & A9:    &  2 & 38.2 &   32&      & in vis. binary, dScut-new           \\
 HD~20790   & 0060-1499 &         &  8.90& F0     & F4     &  1 & 62.4 &   97&      & gDor-new                            \\
            & 0060-0969 &         & 10.20& F8     & K2:    &  1 & 15.7 &     &      & gDor-new                            \\
 HD~20884   & 0060-0509 &         &  7.47& K0     & K1(G)  &  1 & 26.0 &     &      & RGpuls-new                          \\
 HD~21839   & 5296-0253 &         &  9.76& F5     & F5     &  1 &  2.4 &   79&      & var ?                               \\
 HD~21949   & 5296-0283 &         &  9.27& F0     & F4(G?) &  1 &  6.8 &     &      & LPV-new                             \\
 HD~21994   & 5296-0672 &         &  8.62& K0     & K0(G)  &  1 & 36.2 &     &      & RGpuls-new                          \\
 HD~293785  & 4758-0092 &         &  9.29& K7     & K2(G)  &  1 &-44.8 &     &      & LPV-new                             \\
 HD~32704   & 4758-0066 &         &  8.66& G8V    & G6-7   &  7 &      &     & SB2  & spots-new?                          \\
 HD~258857  & 0158-1480 & NGC2244 &  9.02& K2IV-V & K2     &  1 & 31.4 &     &      & LPV-new                             \\
 HD~258916  & 0158-0444 & NGC2244 &  9.64& A5     & F2     &  4 &      &     & SB2  & W UMa-known                         \\
 HD~46105A  & 0158-2674 & NGC2244 &  7.13& A1p    & A1:    &  9 &      &     & SB2  & in vis binary, spots-new            \\
 HD~46105B  & 0158-2674 & NGC2244 &  8.62&        & A7:    &  1 & 28.8 &     &      &                                     \\
 HD~46180A  & 0154-2360 & NGC2244 &  9.10& A3V    & A3:    &  6 &      &     & SB2  & in vis binary, eclBin-new           \\
 HD~46180B  & 0154-2360 & NGC2244 &  9.50&        & F4-5   &  4 &      &     & SB2  & in vis binary, eclBin-new           \\
 HD~46324   & 0154-1047 & NGC2244 &  7.98& K5III  & K2(G)  &  2 & 48.3 &     &      & LPV-new                             \\
 HD~46375   & 0154-0891 & NGC2244 &  7.84& K1IV   & K1     &  2 &  0.8 &     &      &                                     \\
 HD~259696  & 0154-0435 & NGC2244 &  8.85& K7     &        &  1 & 16.7 &     &      & LPV-new                             \\
 HD~46517   & 0154-0867 & NGC2244 &  8.07& A7V    & F0     &  2 & 41.7 &  175&      & dScut-gDor-new                      \\
 HD~261230  & 0750-0863 & NGC2264 &  9.39& A7V    & F2     &  1 & 21.0 &  104&      & dScut/PMS-new?                      \\
 HD~261355  & 0750-1713 & NGC2264 &  8.23& F5-A0  & F3     &  1 &-10.4 &   44&      & PMS-new                             \\
 HD~261387  & 0750-1597 & NGC2264 & 10.60& A2V    & A1     &  1 & -8.8 &  127&      & PMS-new                             \\
 HD~47554   & 0750-1471 & NGC2264 &  7.86& F2V    & F5     &  2 & 34.6 &   31&      & gDor-new                            \\
 HD~261683  & 0746-1733 & NGC2264 &  8.13& K5III  & K3-4(G)&  1 &  8.5 &     &      & RGPuls-new?                         \\
 HD~261783  & 0750-1163 & NGC2264 &  8.28& K3II-III & K3(G)&  1 & 17.3 &     &      & LPV-new                             \\
 HD~262014  & 0750-1525 & NGC2264 &  9.95& A7V    & F1-2   &  1 & 38.6 &   52&      & var-new?                            \\
 HD~262500  & 0750-1227 & NGC2264 &  9.96& A6V    & F2     &  1 & 21.8 &  159&      & dScut-new                           \\
 HD~48331   & 0750-0423 & NGC2264 &  9.07& A0     & F3     &  3 & 16.7 &     &      & spots-new                           \\
 BD+10 1246 & 0750-0455 & NGC2264 &  8.37& M1     &        &  1 &  6.6 &     &      & var-new                             \\
 HD~263236  & 0750-0685 & NGC2264 &  8.10&        & F4     &  1 & 41.2 &   38&      & gDor-new                            \\
 HD~263551  & 0755-0575 &         &  9.68& A7     & F3     &  2 & 23.4 &   67&      & dScut-new                           \\
 HD~58714   & 0764-0548 &         &  6.73& K0     & K2-3(G)&  1 & 56.5 &     &      & LPV-new                             \\
 HD~61199   & 0187-1892 &         &  8.18& A3     & F2     &  2 &      &     & SB2-3& dScut-new                           \\
 BD+05 1735 & 0187-1557 &         &  8.75& M2III  &        &  1 &  1.7 &     &      & LPV-new                             \\
 HD~61696   & 0187-1562 &         &  8.06& K0     & K0(G)  &  1 & 25.2 &     &      & RGPuls-new                          \\
 BD+06 1752 & 0191-1302 &         & 10.16& F2     & F4-5   &  2 & -8.9 &   37&      & dScut-new                           \\
 BD+05 1747 & 0187-1737 &         &  8.98& K2     & K2(G)  &  1 & 26.9 &     &      & RGPuls-new?                         \\
 HD~73709   & 1395-2006 & M44     &  7.31& F2III  & F5:    & 14 &      &     & SB1  & in vis. binary, eclBin-new          \\
 BD+12 1913 & 0813-1617 & M67     &  9.42& K2     & K3(G)  &  1 & 33.1 &     &      & RGPuls-new                          \\
 BD+12 1918 & 0813-1152 & M67     &  8.89& K0     & K0(G)  &  1 & 39.5 &     &      & RGPuls-new                          \\
 BD+12 1919 & 0813-1032 & M67     &  8.67& K5     &        &  1 & 32.5 &     &      & RGPuls-LPV-new                      \\
 BD+28 1653 & 1949-1584 &         &  8.99& K0     & K2(G)  &  1 & 33.4 &     &      &                                     \\
            & 0814-1493 & M67     &  9.68& K2III  & K2(G)  &  1 & 33.1 &     &      & RGPuls??                            \\
            & 0814-2331 & M67     &  8.03& K4III  & K2(G)  &  1 & 34.2 &     &      & RGPuls                              \\
 BD+12 1924 & 0814-2405 & M67     &  9.25& K2     &        &  1 & 35.8 &     &      & RGPuls-new                          \\
            & 0814-0889 & M67     & 11.17& F4     & F4     &  2 &      &     & SB3  & eclBin-known, Algol                 \\
 HD~75638   & 0814-0601 & M67     &  7.76& F0     & F5     &  8 &      &     & SB3  & in vis. binary, eclBin-new          \\
 BD+12 1926 & 0814-1515 & M67     &  9.47& K3III  & K2(G)  &  1 & 34.1 &     &      & RGPuls-new?                         \\
 HD~75700   & 0814-0795 & M67     &  6.73& K0     & K1-2(G)&  2 & -2.0 &     &      & RGPuls-new                          \\
 HD~75717   & 0817-1563 & M67     &  8.63& F2     & F2     &  2 &  7.9 &   67&      & dScut-new                           \\
 BD+12 1929 & 0814-0323 & M67     &  8.88&        & F8-9   & 20 &      &     & SB2  & eclBin-new                          \\
 BD+12 2159 & 0833-1094 &         &  8.98& K2     & K1(G)  &  2 & 13.8 &     &      & RGPuls-new?                         \\
 HD~102083  & 0275-0271 &         &  8.59& A5e    & F3-4   &  1 & 12.1 &  214&      & dScut-known                         \\
 HD~103548  & 0273-0462 &         &  7.15& F2     & F4-5(G)&  2 &  3.9 &   32&      & LPV-new                             \\
 HD~114839  & 1996-1075 &         &  8.46& Am     & F4-5   &  2 & -8.4 &   70&      & dScut-new                           \\
 HD~116462  & 5547-0592 &         &  7.26& F0     & F4     &  3 & -1.8 &   48&      & gDor-new                            \\
 BD+17 2663 & 1460-0878 &         &  9.64& M0     &        &  3 &-10.4 &     &      & LPV-new                             \\
 BD+17 2667 & 1460-0976 &         &  9.95& F8     & F5     &  4 &-17.9 &     &      &                                     \\
 BD+17 2670 & 1460-0532 &         &  9.82& K2     & K2(G)  &  2 &-48.7 &     &      &                                     \\
\hline
\end{tabular}
\end{center}
\end{scriptsize}
\end{table*}

\begin{table*}
\addtocounter{table}{-1}
\begin{scriptsize}
\caption{(continued)}
\begin{center}
\begin{tabular}{llrrllrrrcl}
\hline
  HD/BD     &   GSC     & Cluster &  $V$ & Sp.    & Sp     & \# & RV    &$v \sin i$ & Multiplicity   & Note                    \\
            &           &         &      & publ.  & BF     &    & [km/s]&  [km/s]   &           &                         \\
 BD+18 2797 & 1467-0355 &         &  9.39& M5     &        &  1 & -4.2 &     &      & XZ Boo, LPV-known                   \\
 HD~124569  & 1472-0535 &         &  7.48& M0III  &        &  1 & -5.5 &     &      & LPV-new                             \\
            & 1472-0811 &         &  9.71&        & M3-4(G)&  1 &-36.9 &     &      & LPV-new                             \\
 HD~142536  & 0943-0295 &         &  8.10& K5     &        &  3 & 26.4 &     &      & LPV-new                             \\
 HD~142639  & 0943-0347 &         &  6.89& K0     & K0(G)  &  3 & 10.8 &     &      & LPV-new                             \\
 HD~149758  & 5636-0050 &         &  8.31& K5     &        &  2 &-51.3 &     &      & LPV-new                             \\
 HD~149977  & 5636-0153 &         &  9.02& K5     & K0(G)  &  2 &-39.8 &     &      & LPV-new                             \\
 HD~175922  & 1051-1124 &         &  7.21& Am     & F5     &  1 &-59.7 &     &      & spots-new                           \\
 HD~176285  & 1051-1430 &         &  9.51& F8     & F4:    &  2 &-14.6 &     &      &                                     \\
 HD~176948  & 5136-0659 &         &  8.10& K2     & K3(G)  &  2 &  6.7 &     &      & LPV-known                                 \\
 HD~176949  & 5140-2635 &         &  8.88& G5     & G2(G)  &  2 & 12.4 &     &      & LPV-new                             \\
 BD-05 4861 & 5136-1743 &         & 10.20& F5     & F4-5   &  3 &-26.5 &  112&      & V Algol, no-eclipses                     \\
 HD~345479  & 1628-2407 &         & 11.30& F2     & F3     &  1 &      &     & SB2  & gDor-new                            \\
 HD~345478  & 1628-2345 &         & 10.51& F0     & F4     &  1 & 26.2 &  119&      & dScut-new                           \\
 HD~345467  & 2141-0566 &         &  8.22& M2     &        &  1 & -2.1 &     &      & LPV-new                             \\
 HD~345468  & 2141-0922 &         &  9.55& F0     & F4-5   &  5 &      &     & SB2? & in vis. binary, gDor-new            \\
 HD~345444  & 2141-2482 &         & 10.53& A3     & F2     &  1 &-16.8 &  114&      & dScut-new                           \\
            & 2141-1146 &         &  9.69&        & K0:    &  2 &-26.1 &     &      & LPV-new                             \\
 HD~345455  & 2141-0526 &         & 10.48& B9     &        &  2 & -5.6 &     &      & eclBin-new                          \\
 HD~345459  & 2141-0916 &         &  7.58& K0     & K1(G)  &  3 &  7.8 &     &      & LPV-new                             \\
 BD+08 4628 & 1105-0367 &         &  9.33& K2     &        &  1 &  5.2 &     &      & LPV-new                             \\
 BD+08 4631 & 1109-2546 &         &  9.35& G0     & G8-9   &  1 &-33.7 &     &      & LPV-new                             \\
 BD+18 4914 & 1688-1766 &         & 10.60& F5     & F5-6   &  1 &-38.5 &   40&      & gDor-dScut-known                    \\
 BD+18 4923 & 1688-0773 &         &  8.72& K2     &        &  1 &  9.7 &     &      & LPV-new                             \\
 HD~209775  & 1684-1373 &         &  7.59& F0     & F5-6   &  1 &  1.1 &   85&      & dScut-known                         \\
 HD~217423  & 3216-2213 &         &  8.71& F0     & F0     &  1 & 10.2 &  191&      & dScut-gDor-new                      \\
\hline
\hline
\end{tabular}
\end{center}
\end{scriptsize}
\flushleft{Explanation of columns:
HD/BD/GSC -- catalog identifications; Cluster membership;
$V$ -- visual magnitude; Sp. (publ.) -- published spectral type;
Sp. (BF) -- spectral type estimated from the BF strength;
\# -- number of observed spectra;
RV -- the mean radial velocity;
$v \sin i$ -- projected rotational velocity;
Multiplicity -- as observed;
Photometric variability type according to MOST observations.
The abbreviations used:
dScut -- $\delta$~Sct pulsations, eclBin -- eclipsing binary, gDor -- $\gamma$~Dor pulsations, 
LPV -- Long-Period Variable, PMS - pre-main sequence variable, RGpuls -- pulsating red giant.
Few constant stars were also observed (given without variability type)
\\
\smallskip
Notes on individual systems:\\
HD~965 - inconsistent rotational velocity: while our spectra indicate 
         $v \sin i < 20$ km~s$^{-1}$, \citet{uesu1970} give $v \sin i$ = 90 km~s$^{-1}$\\
HD~9014 - SB2 composed of rapidly rotating F-type component
with $v \sin i$ = 212$\pm$3 km~s$^{-1}$ and slowly rotating much fainter
component ($L_2/L_1 \approx 0.04$) with projected rotational rate below our
spectral resolution\\
GSC~2816-2176 - according to \citet{daniel94} it is SB1 system. Our spectra,
however, show two components of similar brightness\\
HD~32704 -- relative intensity of components in BFs was found to significantly vary\\
HD~46180A -- similar to HD~9014, rapidly rotating component has
$v \sin i$ = 125$\pm$2 km~s$^{-1}$, ratio of intensities of components is
$L_2/L_1 = 0.11\pm0.03$\\
HD~61199 -- our BF is blend of three components, separate RVs not determined\\
GSC~1472-0811 -- TiO absorption strength indicates M3-4III spectral type
\\
\smallskip
GCVS designations: HD~6242 = CQ~Psc, HD~9289 = BW~Cet, HD~258857 = V848~Mon,
GSC~0814-0889 = ES~Cnc, BD+18 2797 = XZ~Boo, HD~209775 = V377~Peg
}
\end{table*}

While not all of some 150 initially selected 
targets have been observed in
this program, we have decided to publish our results because
the David Dunlap Observatory was closed on
2008 July 2 and no longer functions. 
Because we have no access to a similar 2-metre class telescope 
and because the current data form a homogenous and uniform sample,
this survey is published in the present form, as one of the last
legacies of this venerable observatory.

The present paper gives for the survey targets
(i)~an estimate of the spectral type,
(ii)~radial velocity (for most of stars just one observation
is available), (iii)~the projected rotational
velocity, for cases of $v \sin i > 30$ km~s$^{-1}$.
The spectral type was estimated on the basis of the best
template producing the integral 
of the broadening-function (BF) closest 
to unity (see Section~\ref{analysis}).
For a few stars previously known or discovered by us to
be spectroscopic binaries, we present radial velocities for
individual components and preliminary orbits.
MOST photometry of the newly discovered eclipsing binaries
will be presented in a separate paper. Analysis of variables
(including eclipsing binaries) in the field of M67 
is given in \citet{m67}. We note that 
many new variables are $\delta$ Scuti or $\gamma$~Doradus
pulsators, where estimates of the spectral type 
and the projected rotational velocity,
$v \sin i$, are important data for the asteroseismology modelling.

\section{Spectroscopic Observations}
\label{spectroscopy}

Spectroscopic observations were obtained using the slit spectrograph
in the Cassegrain focus of 1.88m telescope of the
David Dunlap Observatory (DDO). The spectra were taken
in a spectral window of about 240~\AA\ around the Mg~I
triplet (5167, 5173 and 5184~\AA) with an effective
resolving power of $R$ = 12,000 -- 14,000. The diffraction
grating with 2160 lines/mm was used with the sampling of
0.117 \AA/pixel.

\begin{table*}
\caption{Parameters of the template stars used to extract broadening
         functions and to define the radial velocity system. HD~107966 was
         regularly observed, but due to its high rotational velocity it
         was not finally used as a template. The source of metallicity
         (always given relative to that of the Sun) is given in the last column.
         \label{template.tab}}
\footnotesize
\begin{center}
\begin{tabular}{rlrccrl}
\hline
\hline
No.&  Name     &   RV   &$v \sin i$ & sp. & [Fe/H] & Ref.              \\
   &           & [km/s] &  [km/s]   &     &        &                   \\
\hline
0 & HD~72660  &   2.8  &  9 & A1V        &   0.34  & \citet{lemk1989}  \\
1 & HD~107966 &   1.3  & 55 & A3V        &         &                   \\
2 & HD~128167 &   0.2  &  5 & F2V        &$-$0.41  & \citet{edvar1993} \\
3 & HD~693    &  14.4  &  8 & F5V        &$-$0.38  & \citet{edvar1993} \\
4 & HD~102870 &   4.6  &  3 & F8.5IV-V   &   0.13  & \citet{edvar1993} \\
5 & HD~114710 &   6.1  &    & G0V        &   0.03  & \citet{edvar1993} \\
6 & HD~23169  &  17.4  &    & G2V        &         &                   \\ 
7 & HD~32923  &  20.3  &    & G4V        &$-$0.20  & \citet{hear1972}  \\
8 & HD~65583  &  13.2  &    & G8V        &$-$0.70  & \citet{soub2008}  \\
9 & HD~185144 &  26.7  &  8 & K0V        &$-$0.21  & \citet{soub2008}  \\
10& HD~3765   &$-$62.3 &    & K2V        &   0.10  & \citet{soub2008}  \\
11& HD~160964 &$-$28.4 &    & K4V        &         &                   \\
12& HD~79210  &  11.1  &    & M0V        &         &                   \\
\hline
\hline
\end{tabular}
\end{center}
\end{table*}

The region of the magnesium triplet has been
used throughout the extensive short-period binary program
(see \citet{ddo13} for the most recently published
paper of the DDO series) and we continued using it for this survey.
The Mg~I triplet feature serves exceedingly well for
majority of late spectral types. Because of size constraints,
short-period binaries with orbital periods shorter than
one day which were observed during the binary
program were practically all of late spectral types.
For stars earlier than A0, however, the magnesium triplet
becomes very weak and practically unusable for extraction
of any radial velocity and broadening information.
Even for mid-A type stars, metallic lines
are weak and a high S/N is required. Therefore, several
spectra of stars earlier than about A1 were found to be unusable
and thus discarded from this survey. Three targets,
HD~31919, HD~32884 and HD~45972, which were included
in the survey and assumed to have sufficiently late spectral
types for the Mg~I observations, were found later to be too
early; they have been omitted from the tables.

One-dimensional spectra were extracted by the usual
procedures within the IRAF environment\footnote{IRAF
is distributed by the National Optical Astronomy
Observatories, which are operated by the Association
of Universities for Research in Astronomy, Inc., under a
cooperative agreement with the NSF.} after the bias and
flat-field photometric correction of the original
frames. Cosmic-ray trails were removed using a
program of \citet{pych2004}.

\begin{figure*}
\includegraphics[width=170mm,clip=]{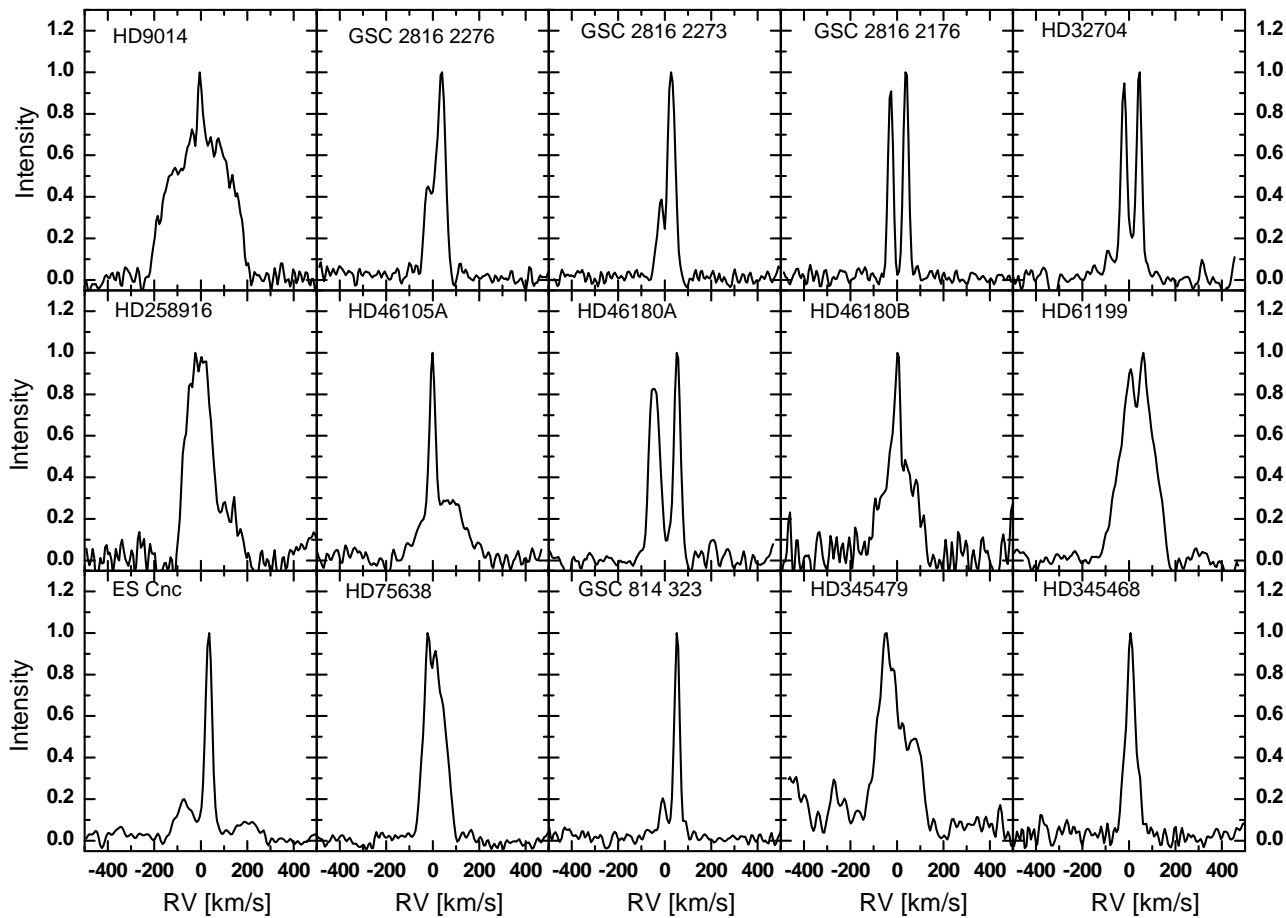}
\caption{Selected broadening functions of SB2 and triple systems observed
         at DDO.
\label{bfplots}}
\end{figure*}

The current survey was limited to objects north of
$-15 \degr$, and stars brighter that $V = 11.5$. The data were obtained
between 2007 November 13 and 2008 June 19. Because of the
observatory closure,
most of the targets were observed only once or twice and
the goals of the survey were only partially fulfilled.
In total, 241 spectra of 103 MOST targets (96 of them being
photometrically variable) were obtained. An overview of the systems
and their observations is given in Table~\ref{main.tab}. The dates of all
spectroscopic observations are given in Table~\ref{rv.tab} listing
the mean observed RVs for single and multiple stars, respectively.
Selected BFs of spectroscopic binaries and triple systems
are shown in Fig.~\ref{bfplots}.

\begin{table*}
\caption{Radial velocities and their standard errors for all single stars, 
SB1 and SB2 systems and triple system ES~Cnc (the full table is available 
only in electronic form, here RVs are given for first three SB2 systems and 
one single star). The column denoted as ``Temp.'' gives the template number 
used to extract BF and determine the RVs referring, as in the first column 
of Table~\ref{template.tab}. The components were designated according to 
observed relative brightness with the component 1 assumed being always the 
brighter one. The RV for SB2 systems is given if the components could be 
separated only. Type of the function to fit to BFs is given either as 
G (Gaussian) or R (rotational profile) in the columns denoted ``Fit''. 
Radial velocity is not given for observations of SB2 or triple systems 
leading to inseparable peaks in BFs. \label{rv.tab}}
\footnotesize
\begin{center}
\begin{tabular}{lcrccrcc}
\hline
\hline
  HJD       & Temp. &  RV$_1$ & $\sigma_1$ &  Fit &  RV$_2$  & $\sigma_2$ & Fit \\
2\,400\,000+&       & [km/s]  &  [km/s]    &      & [km/s]   &  [km/s]    &     \\
\hline
{\bf HD 9014}&  &          &      &   &          &      &   \\
54422.60716 & 2 &  $-2.81$ & 1.49 & R & $-3.11$  & 1.05 & G \\
54490.46861 & 2 & $-11.73$ & 1.58 & R & $21.96$  & 1.24 & G \\
54494.46390 & 2 & $-14.78$ & 1.71 & R & $23.41$  & 1.11 & G \\
54504.47860 & 2 & $-17.03$ & 2.03 & R & $26.78$  & 1.24 & G \\
{\bf GSC 2816~2276} & &    &      &   &          &      &   \\
54422.65196 & 3 &  $36.86$ & 0.58 & G & $-15.42$ & 1.44 & G \\
54481.66400 & 3 &  $34.28$ & 1.29 & G & $-14.58$ & 2.76 & G \\
54497.51571 & 3 &  $-8.72$ & 1.47 & G &   30.66  & 5.15 & G \\
{\bf GSC 2816~2273} & &    &      &   &          &      &   \\
54418.63188 & 3 &  $-6.61$ & 0.27 & G &   55.07  & 1.69 & G \\
54490.49523 & 3 &  $28.42$ & 0.28 & G & $-20.29$ & 0.73 & G \\
{\bf HD~965}&   &          &      &   &          &      &   \\
54418.50574 & 2 &    0.77  & 0.39 & G &          &      &   \\
54418.53590 & 2 &    0.59  & 0.38 & G &          &      &   \\
54418.54755 & 2 &    0.82  & 0.39 & G &          &      &   \\
\hline
\hline
\end{tabular}
\end{center}
\end{table*}

In addition to the program stars, several
template spectra were observed (see Table~\ref{template.tab})
to be used in the broadening function determination
(see the next Section~\ref{analysis}).
The templates were selected to be of stars with
low projected rotational velocities (especially
for A and F spectral types), known spectral types and
constant radial velocities (hereafter RVs).
Our template sequence is sufficiently continuous
for the broadening function applications except for
a gap between A1V (HD~72660) and F2V (HD~128167),
where we could not find
any suitable templates having a low projected RV.
Several RV standards were also observed
to ascertain their reliability within our RV system.

\section{Analysis of spectra}
\label{analysis}

The observations were analyzed using the technique of broadening-functions (BF),
described in \citet{Rci1992,Rci2002}, using spectra of slowly rotating stars
of similar spectral type as templates. The directly extracted BFs were smoothed 
by a convolution with the Gaussian function to match the spectral resolution of
the spectrograph. Smoothing with $\sigma =1.5$ pixels (10 km~s$^{-1}$) was found 
to be appropriate for most BFs; a few BFs which were very noisy were smoothed 
using $\sigma = 2$ pixels.

Because the observing time available for the observations was rather limited,
mainly due to poor weather at DDO during the time of the survey and -- later -- 
due to the observatory closure, no low-resolution classification spectra were 
taken for direct spectral type and luminosity class estimates, as had been 
originally planned. The spectral types were, however, determined by finding 
a template (listed in Table~\ref{rv.tab}) matching the observed spectrum best. 
For the best fitting template the integrated BF strength is expected to be unity. 
This rough spectral-type estimate is possible because the wavelength range 
contains only metallic lines and their strength is increasing with decreasing 
temperature to about K5 spectral type where molecular bands start to be prominent. 
Therefore using progressively later templates for a given spectrum results in 
decreasing BF integral. For few stars we either could not achieve a good template 
match to the spectrum or the BF strength was not decreasing monotonically with 
progressively later-type templates. This typically occurred for stars later 
than K2--5 or B and A stars (some of which are probably chemically peculiar).
For those stars, the spectral-type estimate based on the BF strength is not 
given (see Table~\ref{main.tab}). Except for the dwarf templates, we observed 
a few giants covering the range of spectral types of G8III to M3III. We noticed 
that using a dwarf template for the spectrum of a giant results in a significant 
depression around the resulting BF base. This could be used for a very approximate 
guess at the luminosity class. For targets showing such a depression which strongly 
indicates a luminosity class II--III, the spectral type in Table \ref{main.tab} 
was augmented by ``(G)'' (giant). In any case the spectral-type estimates should 
be regarded as very approximate because the BF strength is expected to depend 
not only on the temperature but also on the metallicity. 

Determination of $v \sin i$ for low projected
rotational velocities is limited by the
spectral resolution ($R = 12\,000 - 15\,000$) setting a lower limit at about
15 -- 20 km~s$^{-1}$. The Gaussian profiles were fitted to BFs of slowly 
rotating stars with $v \sin i < 30$ km~s$^{-1}$ to determine RV; for those 
stars only the RV and estimated spectral type is given. For rapidly-rotating
stars, the RVs and projected rotational velocities, $v \sin i$,
were determined by fitting rotational profiles to the
extracted broadening functions, as described in
\citet{ddo11}. Because the shape of the theoretical rotational
profile depends (weakly) on the limb darkening in the
5184 \AA\ spectral window, the limb darkening coefficient
was adopted from extensive tables of
\citet{hamme93} according to the estimated spectral type.

Precision of the RV measurements is defined by random errors (influenced
mainly by the S/N ratio of spectra, width and intensity of components in the
extracted BFs) and by systematic effects of spectrograph flexure which are
rather hard to quantify. Our observations did not utilize a iodine cell
and telluric spectral lines were not present in our spectral window so no
simple, external checks on the RV zero-point were available. We attempted
to assess the stability of our RV system by using multiple observations of
bright standard stars. For such stars having at least 5 spectra, the first
spectrum of a given star was used as a template and the RVs were then
determined fitting the (obviously very narrow) Gaussian function to the
extracted BFs for the remaining spectra. As expected, the scatter within
such samples depends on the spectral type of the standard and decreases
for late-type standards with large numbers of strong lines in their
spectra. Along the spectral sequence, we obtained: HD~97633 -- A2V,
$\sigma = 1.13$ km~s$^{-1}$, HD~128167 -- F2V, $\sigma = 0.91$
km~s$^{-1}$, HD~102870 -- F8.5IV-V, $\sigma = 0.84$ km~s$^{-1}$, and HD~144579 -- G8V,
$\sigma = 0.59$ km~s$^{-1}$. Hence, one can expect systematic
deviations of individual RVs to be at the level of typically less than
about 1.0 -- 1.2 km~s$^{-1}$. This effectively means that for measurements 
with formal standard errors given in Table~\ref{rv.tab} smaller than 
about 0.5 -- 1.0 km~s$^{-1}$, the systematic errors dominate in the 
error budget. 

The resulting RVs for most observed systems are given in
Table~\ref{rv.tab} (not given for binaries and multiples
with complicated blends of components).
With many targets observed just once,
we cannot exclude a possibility that we missed several SB1 systems or
SB2 systems observed close to their spectroscopic conjunctions.

\section{Binary and multiple systems}

Spectroscopic observations showed that part of the targets are 
double or triple systems (see Fig.~\ref{bfplots}). RVs of individual 
components in these cases were determined by the same procedures as 
described in \citet{ddo11}. In the case of SB2 systems, double rotational 
or Gaussian profiles
were fitted to the components, depending on the
rotational broadening of the lines; the fitted functions are
indicated in Table~\ref{rv.tab}. For the triple system
ES~Cnc, the third (slowly-rotating) component was first removed after a
multi-component Gaussian model was fit to the observed BFs; 
the profiles of the binary components were then
approximated by rotational profiles.
For the triple systems HD~61199 and HD~75638,
where components are always blended, no individual RVs could be determined.

The present survey has led to detection of 7 new spectroscopic binaries: HD~9014
(GSC~5276-0313), HD~32704 (GSC~4758-0066), HD~46105 (GSC~0158-2674),
HD~46180 (GSC~0154-2360) where both visual components are SB2, GSC~0814-0323,
HD~345468 (GSC~2141-0922; components always blended), and HD~345479 (GSC~1628-2407; 
a marginal detection). Prior to the present spectroscopic survey, MOST photometry 
showed that HD~46180, and GSC~0814-0323 contain eclipsing binaries.

All spectroscopic binaries in NGC~752 observed within this survey were already
detected by \citet{daniel94}. This applies to GSC~2816-2068, GSC~2816-2176,
GSC~2816-2273, and GSC~2816-2276. Unfortunately, the authors did not give any
details regarding the orbital periods or preliminary spectroscopic orbits. The
most interesting and important is GSC~2816-2068, which was found to be
an eclipsing
binary with $P$ = 15.52 days with eclipses about 0.11 mag deep. Four systems,
GSC~0814-0323, HD~73709, HD~46180, and HD~75638 deserve special attention (see the
following subsections). Notes for other systems are given below the main table
(Table~\ref{main.tab}).

The spectroscopic elements (case of GSC~814-323 and HD~73709) were determined by the
differential corrections optimization.\footnote{Unpublished code {\it ELEMDR} of
the first author}

\subsection{GSC~0814-0323}

The only newly discovered SB2 (and a eclipsing binary
discovered by MOST) for which we have enough
data to obtain a preliminary spectroscopic orbit is 
GSC~0814-0323, located in the sky close to the open cluster M67.
The ratio of component luminosities, as estimated from the
BFs when the components are split, is about $L_2/L_1 \sim 0.29$.
With orbital period as long as 68 days, and with the obvious
presence of eclipses, the orbital-plane inclination must
be very close to $90\degr$ so that the projected masses of components,
$M_1 = 1.39 \pm 0.08$ M$_\odot$ and
$M_1$ = 0.96$\pm$0.06 M$_\odot$ must be close to the true ones.
The predicted spectroscopic
conjunction, HJD $2\,453\,547.35\pm0.86$, occurs 8 days
earlier than predicted by an eclipse observed by MOST
and using the spectroscopic period.
There is a single radial velocity measurement of GSC~0814-0323 at
HJD 2\,441\,321.88 published by \citet{math1986}, giving
$RV = 12.6 \pm 0.5$ km~s$^{-1}$.
The measurement probably refers to the brighter component of
the visual pair and hence to the primary component of the
eclipsing system. The light curve of GSC~0814-0323
was discussed in a paper dedicated to MOST photometry of M67
field \citep{m67}.

\subsection{HD~73709}

\begin{figure}
\includegraphics[width=80mm,clip=]{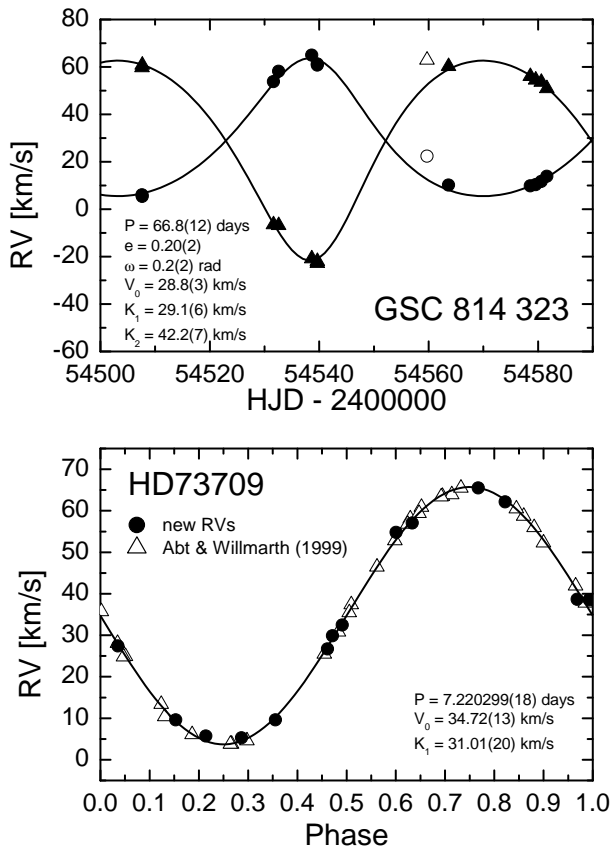}
\caption{The radial velocity orbits for the
SB2 system GSC~0814-0323 (top) and the SB1 system
HD~73709 (bottom). One radial-velocity determination for
the former star (plotted by an open symbols) was not used
in the computation of the orbit. The new RVs of HD~73709
were combined with those published in \citet{abt1999} to refine the orbital
period (see the text). The resulting preliminary orbital elements are
listed in the respective panels. 
\label{orbits}}
\end{figure}

HD~73709 was found to be SB1 by \citet{abt1999} who
determined the following orbital elements:
$P = 7.2204(1)$ days, $V_0 = 34.6(1)$ km~s$^{-1}$ and $K = 31.2(2)$
km~s$^{-1}$, based on 30 RV measurements over 3 years.
MOST photometry showed that the star is an eclipsing binary
with $P = 7.22$ days, which is compatible with the
spectroscopic determination. The system is a high-probability member
of the Praesepe cluster. Our new RVs were combined
with those of \citet{abt1999} to refine the orbital elements.
Since the orbital eccentricity was found to be insignificant,
we assumed the circular orbit.
All RVs and the combined solution are shown in Fig.~\ref{orbits}.

\subsection{HD~46180}

HD~46180 (GSC~0154-2360) is close visual pair in NGC 2244 composed of
components of $V = 9.1 + 9.3$
at $\theta = 82\degr$, and $\rho = 5.2''$ \citep{wds}.
The stars were oriented almost
along the spectrograph slit which enabled us
to simultaneously obtain spectra of either of the components on nights
with the best seeing. Our spectroscopy shows that each of the visual
components is a binary star so that HD~46180 is a quadruple system.
Unfortunately, we were able to obtain only 10 spectra of the brighter
visual component and 4 of the fainter one (on other nights it was
blended with the brighter component), which has prevented us to
determine the respective orbital periods.
MOST satellite photometry refers to the whole
quadruple system and shows that one of the binaries is an eclipsing
one with the period of $P = 3.09$ days and the
photometric amplitude of only 0.02 mag. The present
data indicate that the brighter component of the visual pair is the
variable one.

\subsection{HD~75638}
\label{hd75638}

HD~75638 (GSC~0814-0601) is a visual triple system WDS 08515+1208
consisting of components A~($V = 8.28$), B~($V = 10.17$) and C~($V = 11.8$).
Our spectroscopy refers to the pair AB, $\rho$ = 1.4 arcsec, which
could not be separated on the DDO spectrograph slit. The pair
AB was suspected to
be an SB3 system in the survey of \citet{nord1997}. 
Their analysis of 13 spectra
showed two measurable components with $v_1 \sin i$ = 80 km~s$^{-1}$ and
$v_2 \sin i$ = 10 km~s$^{-1}$, temperatures $T_1 = 7000 K$, and $T_2 = 6750 K$,
and a luminosity ratio of 0.10. A preliminary orbit for the fainter component
was given: $P$ = 5.8167(9) days, $V_0$ = 11.05(91) km~s$^{-1}$,
$K_1$ = 28.9(12) km~s$^{-1}$, $e$ = 0.083(56), and $\omega$ = 322(28)\degr.
The spectroscopic orbit most likely corresponds to the motion of the blend
of a faint pair seen as component B.

The MOST photometry showed very shallow ($\Delta m = 0.013$) eclipses giving a
preliminary mid-eclipse ephemeris, HJD $2\,454\,145.85 + 5.81 \times E$,
in good accord the spectroscopic result of \citet{nord1997} (see \citet{m67}).
Our spectroscopic results are related to the AB
components. The broadening functions
(Fig.~\ref{bfplots}) show the brighter, rapidly-rotating component
($v \sin i \approx 85$ km~s$^{-1}$) to be stationary in the radial velocity;
this star is most likely the A component of the visual pair. The fainter
component B occasionally splits into two.
All three components are always blended,
suggesting that a high-resolution spectroscopy will be necessary to
reliably define orbits of both components.

\section{Conclusions}
\label{sum}

The medium dispersion spectroscopic survey of the
variable stars discovered by the MOST satellite was planned
to include some 150 targets observed several times to
fully ascertain types of variability. This has not been
possible because of the DDO observatory closure.
We publish here radial velocities and spectral type
estimates for 103 MOST targets, 96 being variables with
83 of them being new MOST detections. For 31 targets, we
give the projected rotational velocities $v \sin i$.

The survey resulted in the discovery
of seven new spectroscopic binaries.
For two of these binaries (HD~73709, and GSC~0814-0323), which were
found to show eclipses by the MOST satellite photometry, preliminary
spectroscopic orbits were determined. HD~46180 was found to be composed
of two binary stars forming, very probably, a physical quadruple system.
According to the MOST photometry, one of the binaries is eclipsing.
The newly discovered spectroscopic binaries
will require further spectroscopy to refine their orbital periods
and the remaining orbital parameters.

\section*{Acknowledgments}
This study has been funded by the Canadian Space Agency Space Enhancement
Program (SSEP) with TP holding a Post-Doctoral Fellowship position at the
University of Toronto. The Natural Sciences and Engineering Research Council
of Canada (NSERC) supports the research of SMR. Support from the Polish
Science Committee (KBN grants PO3D~006~22 and P03D~003~24) to WO
is acknowledged with gratitude.

The authors would like to thank the telescope
operators at DDO, Heide de Bond, Toomas Karmo and Archie de Ridder,
for their help. 

The research made use of the SIMBAD database, operated at the CDS,
Strasbourg, France and accessible through the Canadian
Astronomy Data Centre, which is operated by the Herzberg Institute of
Astrophysics, National Research Council of Canada.
This research made also use of the Washington Double Star (WDS)
Catalog maintained at the U.S. Naval Observatory.

\label{lastpage}
\end{document}